# Optical mode crossings and the low temperature anomalies of SrTiO$_3$

E. Courtens,[1] B. Hehlen,[1] E. Farhi,[1] and A.K. Tagantsev[2]

[1] Laboratoire des Verres, UMR 5587 CNRS, Université de Montpellier 2, F-34095 Montpellier, France
[2] Laboratoire de Céramique, EPFL, CH-1015 Lausanne, Switzerland

*Abstract*

Optical mode crossing is not a plausible explanation for the new broad Brillouin doublet nor for the strong acoustic anomalies observed at low temperatures in SrTiO$_3$. Data presented to support that explanation are also inconclusive.

Incipient ferroelectric perovskites, $SrTiO_3$ and $KTaO_3$, exhibit low temperature ($T$) anomalies that attract current attention. It was recently proposed that the anomalies in $SrTiO_3$ can all be explained by the accidental crossings of the soft structural modes ($A_{1g}$ and $E_g$) with the soft polar ones ($A_{2u}$ and $E_u$) [1,2]. In their seminal paper, Müller *et al.* [3] already remarked that the lowest crossing occurs near $T_q$ 37 K ($E_g$ with $E_u$), and that below $T_q$ the polarization fluctuations are slower than structural ones. One of us also pointed out [4], independently from [1], that the anharmonic mixing of modes of different symmetries, probably enhanced by fluctuating paraelectric clusters, may explain violations of selection rules in neutron scattering spectra [5,6]. However, the presentation in [2] mixes up several issues: the broad Brillouin doublet (BBD) [7], the indications for a possible transition at $T_q$ [3,4], the unexpected components in neutron spectra [5,6], and the anomalies seen in Brillouin velocities [6,8]. As the question is not resolved, all suggestions are worth considering. It is in this spirit that the proposal [1,2] is discussed here. We find that mode crossing (MC) is irrelevant to the BBD and to the anomalies observed in Brillouin scattering from transverse acoustic (TA) modes.

The BBD of $SrTiO_3$ appears at $T$ well above $T_q$ and remains down to $T \ll T_q$ [6,7]. A similar feature is seen in $KTaO_3$ [6,7] which stays cubic and exhibits no soft structural mode. The BBD is characterized by an "acoustic-like" dispersion, $\Omega = Wq'$. Here, $\Omega$ is the doublet frequency shift, $q'$ is the scattering vector in the notation used in [2], and W is a velocity. It is stated in [2] that two-phonon difference scattering (TPDS) from crossing optical branches, where the two phonons belong to the *different* branches, produces the BBD. In general, TPDS from different branches hardly gives an acoustic-like signature owing to simple kinematic reasons, whereas TPDS from excitations belonging to *one and the same* phonon sheet leads with no difficulty to such a dispersion. There is also a serious intensity problem associated with the MC-proposal: the scattering from this higher order process is expected to be a few orders of magnitude smaller than TPDS from a single sheet. In the latter case, the contribution to the optical dielectric constant ε that controls light scattering is, *e.g.* for TPDS from an acoustic sheet, proportional to $\tilde{u}\tilde{u}$, where $\tilde{u}$ is the acoustic deformation tensor. In the former case, the symmetry imposes that contributions to ε arise from terms involving $\vec{\nabla}\tilde{P}\vec{\tilde{\phi}}_0\vec{\tilde{\phi}}$, where $\tilde{P}$ is the polarization amplitude, $\tilde{\phi}$ the structural mode amplitude, and $\tilde{\phi}_0$ the structural order parameter. The presence of both the gradient and of $\tilde{\phi}_0$ leads to much smaller intensities in this case. To save the situation, [2] invokes a resonance at exact crossing. Exact crossing also gives the required linear dispersion, however with $W \propto q$, the wavevector at the exact crossing point of the optical modes. This wavevector is *strongly T*-dependent, and thus this might lead *at best* to a very rapid variation of W with $T$. Finally, an important condition for TPDS to produce a doublet is that the difference $\Delta = \Omega_1 - \Omega_2$ be sufficiently well defined, *i.e.* that the combined widths of the modes at $\Omega_1$ and $\Omega_2$ be narrower than $\Delta$. From this discussion we identify three experimental facts, each invalidating the explanation proposed in [2]:

(*i*)  a BBD is observed in $KTaO_3$ where there is no mode crossing;

(*ii*) the doublet in SrTiO$_3$ is observed with near constant W over a wide *T*-interval [6,7], contrary to the predictions of the MC-model;

(*iii*) the width of the $E_u$ mode is known near zone center [9], where it is already broader than the BBD frequencies **Error! Bookmark not defined.**, so that $E_u$ cannot produce a real doublet in TPDS.

Recently we re-investigated KTaO$_3$ in detail. We find that the BBD, whether high-frequency second sound (HFSS) [10] or TPDS, is dominated in that case by narrow valleys in the TA sheets [11], rather than by optic sheets. It should be noted that for either TPDS or HFSS to produce a doublet, it is necessary that the thermal-phonon *resistive* scattering rate, **Error! Bookmark not defined.**$_R$**Error! Bookmark not defined.**[1], be smaller than **Error! Bookmark not defined.**. Hence an argument sometimes raised against HFSS ("second sound is impossible in such imperfect materials") were, if valid, equally usable against TPDS for the reasons explained above. For genuine HFSS, the *normal* relaxation rate, **Error! Bookmark not defined.**$_N$**Error! Bookmark not defined.**[1], must in addition be larger than **Error! Bookmark not defined.**, which appears possible in incipient ferroelectrics [10].

Ref. [2] also claims that the non-linear interaction of acoustic modes with the crossing optical modes produces sizeable elastic anomalies at $T_q$. In fact, by far the strongest low-*T* anomaly observed in Brillouin scattering [8] is *not* at $T_q$, but at *T* significantly smaller than $T_q$, ~25 K and lower. Weak anomalies might occur near $T_q$ [6] which, according to our estimates, are of the order of the classical fluctuation corrections to the elastic moduli [12]. The effects invoked in [2] are a generalization of these corrections, and although they are possible, they are expected to be much weaker. This can be shown by comparison of the free-energy invariants responsible for these effects. In the classical case these invariants involve products like $\breve{u}\tilde{\varphi}\tilde{\varphi}$ and $\breve{u}\tilde{P}\tilde{P}$. In the MC-case, simple symmetry arguments lead to terms of the type $\overset{t\pm}{u}\overset{\pm}{\nabla}\overset{r}{P}\overset{r}{\varphi}_0\tilde{\varphi}$, which for the same reason as in the discussion of the BBD above, give much weaker corrections. All in all, the non-linear interaction of acoustic modes with crossing optical modes is expected to be too weak to explain the *strong* TA-anomaly, which also occurs at $T \ll T_q$.

In support of his ideas, Scott also shows [2] ultrasonic data [13] obtained on SrTiO$_3$ that had not been forced into a single tetragonal orientation [14] below the structural transition at $T_a$ **Error! Bookmark not defined.** 108 K. The velocities obtained in that case combine significantly different elastic constants, like $C_{44}$ and $C_{66}$ which have *opposite T*-dependencies. Our opinion is that it is not reasonable to discuss these results until confirmed by experiments on oriented tetragonal samples. In fact it is already known that forcing a **Error! Bookmark not defined.**0 0 1**Error! Bookmark not defined.** SrTiO$_3$ bar into a single c-axis orientation, and measuring the ultrasonic shear velocity along the bar, a perfectly smooth *T*-dependence below $T_a$ is obtained [15], as illustrated in Fig. 2 of [8].

We conclude that mode crossing is unrelated to some of the main low temperature anomalies of SrTiO$_3$.